\documentclass[twocolumn, english, prd,superscriptaddress,
aps, showpacs, showkeys, amsmath, amssymb, floatfix, nofootinbib]{revtex4-1}

\pdfoutput=1
\usepackage[pdftex]{graphicx}
\usepackage[pdftex]{color}
\usepackage[colorlinks,bookmarks]{hyperref}
\usepackage{multirow}
\usepackage[export]{adjustbox}
\definecolor{linkblue}{rgb}{0,0,0.8}
\definecolor{linkgreen}{rgb}{0,0.5,0}
\hypersetup{linkcolor=linkblue, citecolor=linkgreen, urlcolor=linkblue}

\usepackage[T1]{fontenc}
\usepackage[utf8]{inputenc}
\usepackage{lmodern}
\usepackage{babel}
\usepackage{bm}
\usepackage{esint}
\usepackage{verbatim}
\usepackage[us]{datetime}
\usepackage{booktabs}
\usepackage{enumerate}
\usepackage{graphics}
\usepackage{float}
\usepackage{amsmath}
\usepackage{amsfonts}
\usepackage{amssymb}
\usepackage[export]{adjustbox}
\usepackage[normalem]{ulem}
\graphicspath{{./figs/}}

\definecolor{valecol}{rgb}{0,0.5, 1.}

\begin{document}
\title{A model-independent reconstruction of dark sector interactions}

\author{Rodrigo~von~Marttens}
\affiliation{Observat\'orio Nacional, 20921-400, Rio de Janeiro, RJ, Brasil}
\author{Javier E. Gonzalez}
\affiliation{Facultad de Ciencias e Ingeniería, Universidad de Manizales, 170002, Manizales, Colombia}
\affiliation{Departamento de Física, Universidade Federal do Rio Grande do Norte, 59072-970, Natal, RN, Brasil}
\author{Jailson Alcaniz}
\affiliation{Observat\'orio Nacional, 20921-400, Rio de Janeiro, RJ, Brasil}
\author{Valerio Marra}
\affiliation{Núcleo de Astrofísica e Cosmologia (Cosmo-ufes) \& Departamento de Física, Universidade Federal do Espírito Santo, 29075-910, ES, Brasil}
\affiliation{INAF - Osservatorio Astronomico di Trieste, via Tiepolo 11, 34131 Trieste, Italy}
\affiliation{IFPU -- Institute for Fundamental Physics of the Universe, via Beirut 2, 34151, Trieste, Italy}
\author{Luciano Casarini}
\affiliation{Departamento de Física, Universidade Federal de Sergipe, 49000-000 São Cristóvão, SE, Brasil} 
\affiliation{Institute of Theoretical Astrophysics, University of Oslo, P.O. Box 1029 Blindern, N-0315 Oslo, Norway}

\begin{abstract}
Relaxing the conventional assumption of a minimal coupling between the dark matter (DM) and dark energy (DE) fields introduces significant changes in the predicted evolution of the Universe. Therefore, testing such a possibility constitutes an essential task not only for cosmology but also for fundamental physics. In a previous communication [Phys. Rev. D99, 043521, 2019], we proposed a new null test for the $\Lambda$CDM model based on the time dependence of the ratio between the DM and DE energy densities which is also able to detect potential signatures of interaction between the dark components. In this work, we extend that analysis avoiding the $ \Lambda$CDM assumption and reconstruct the interaction in the dark sector in a fully model-independent way using data from type Ia supernovae, cosmic chronometers and baryonic acoustic oscillations. According to our analysis, the $\Lambda$CDM model is consistent with our model-independent approach at least at $3\sigma$ CL over the entire range of redshift studied. On the other hand, our analysis shows that the current background data do not allow us to rule out the existence of an interaction in the dark sector. 
Finally, we present a forecast for next-generation LSS surveys. In particular, we show that Euclid and SKA will be able to distinguish interacting models with about 4\% of precision at $z\approx 1$.
\end{abstract}

\keywords{Cosmology: theory -- dark energy -- dark matter -- large-scale structure}

\date{{\today}}
\maketitle

\section{Introduction}

The standard $\Lambda$-cold dark matter ($\Lambda$CDM) model provides a remarkably successful description of the universe on large scales~\cite{Aghanim:2018eyx,Abbott:2018wzc,Abbott:2018xao}. In this model, the dark energy and dark matter components are minimally coupled to each other and dominate the structure and evolution of the universe in late times. Recently, the validity of the $\Lambda$CDM model has been questioned given the discrepancies and tensions between  early and the late time Universe measurements~\cite{Verde:2019ivm,Riess:2020sih,Asgari:2020wuj}, and a possible alternative could be models with a non-minimally coupling between dark energy and dark matter~\cite{Zimdahl:2001ar,Alcaniz:2002fy}.

As it is well known, there is no known fundamental principle that prevents a 
coupling term $Q$ between the energy components of the cosmological dark sector. Physically, $Q$ represents an energy exchange between the dark components, which necessarily violates adiabaticity at the same time that it brings about important consequences on the model predictions. Currently, a number of analysis show that some particular  classes of interacting models are able to provide a good description of the data~\cite{vonMarttens:2018iav,Benetti:2019lxu,Xia:2016vnp,Kumar:2017dnp,Yang:2019uzo,Martinelli:2019dau,Cid:2018ugy,Yang:2017ccc,Marttens:2017njo,Marttens:2014yja,daSilva:2020ylz,Gonzalez:2018rop}, sometimes alleviating some of the tensions of the standard cosmology~\cite{DiValentino:2017iww,Yang:2018uae,Kumar:2019wfs,Pan:2019jqh,DiValentino:2019jae,Yang:2018euj,Pan:2019gop,Pan:2020bur}. From the theoretical point of view, however, critiques to these models do exist, being mainly related to the absence of a natural guidance from fundamental physics on the coupling term, which leads to a phenomenological choice of $Q$~\cite{Valiviita:2008iv,Costa:2009wv,Cai:2009ht,Carneiro:2019rly}.

In this paper, instead of assuming a parameterization of $Q$ a priori, we take a different route and reconstruct physical quantities directly related to the coupling term  
 from observational data. Different reconstruction methods have been widely used in Cosmology in different contexts~\cite{Zhao:2017cud,Joudaki:2017zhq,LHuillier:2017ani,Shafieloo:2018gin,LHuillier:2018rsv,Keeley:2019esp,Keeley:2019hmw,Liao:2020zko,Keeley:2020aym}. The method we adopt in this work is the \textit{Gaussian Process}~\cite{Seikel:2012uu}.
 
Extending previous results~\cite{vonMarttens:2018bvz}, we employ a reconstruction method to map the evolution of the interaction in the dark sector over a large range of redshift in a model-independent way using data from type Ia supernovae, cosmic chronometers and baryonic acoustic oscillations. Our results show a good agreement with the standard cosmology ($Q=0$), although the existence of an interaction in the dark sector cannot be ruled out. For completeness, we also perform a forecast analysis for some next-generation galaxy surveys and discuss their ability to constrain the possibility of a non-minimally coupling between the dark components.

This work is outlined as follows: In Sec.~\ref{sec.dark} we present a general framework for describing the cosmological background with an unified dark sector. In particular we discuss how the two important features affect the dynamics of the dark sector: the dynamical character of the DE component and the possibility of the dark components interact with each other. In Sec.~\ref{sec.interacting} we develop the formalism for depicting a general interacting dark sector. A slight assumption that the dark sector interaction depends on the energy of the involved components is made, but no model is specified. In Sec.~\ref{sec.reconst} we finally examine the main point of the paper: how to perform a model-independent reconstruction of a dark sector interaction. We develop the relevant equations for the reconstruction, analyze the data employed and present the results obtained from $H\left(z\right)$ measurements (Cosmic Chronometers and BAO) and type Ia SNe (Pantheon). Sec.~\ref{sec.futperspec} is focused to shed a light on how future surveys will be able to improve the results of the proposed analysis. In this analysis we use forecasts for measurements of $H\left(z\right)$ from J-PAS~\cite{Benitez:2014ibt,Bonoli:2020ciz}, DESI~\cite{Aghamousa:2016zmz}, Euclid~\cite{Amendola:2012ys} and SKA~\cite{Bacon:2018dui}. Lastly, Sec.~\ref{sec.conclusions} is dedicated to the concluding remarks.

\section{Unified dark sector: background description}
\label{sec.dark}

Let us consider the FLRW line element ($c = 1$)
\begin{equation} \label{flrw}
ds^{2}=dt^{2}-a^{2}\left(t\right)\left[dr^{2}+r^{2}\left(d\theta^{2}+\sin^{2}\theta d\phi\right)\right]\,,
\end{equation}
where $a$ is the scale factor. For the material content, we assume a Universe composed of four components: radiation (denoted by the subindex $r$), baryons ($b$), cold dark matter ($c$) and dark energy ($x$). At the background level, each one of these components is described as a fluid with an equation of state (EoS) $p_{i}=w_{i}\rho_{i}$, and while the radiation and baryonic  components are separately conserved, DM is allowed to interact with DE. Initially, we do not assume any particular parameterization for the DE component, which means that $w_{x}$ is allowed to be a function of the scale factor. As will be discussed later, we will also consider the specific case $w_{x}=-1$, which, in the absence of interaction, is equivalent to a cosmological constant $\Lambda$. 


Within the general relativistic framework and assuming spatial flatness, the background dynamics is given by the Friedmann equations,
\begin{subequations}
\begin{eqnarray}
    \left(\dfrac{\dot{a}}{a}\right)^{2}\equiv H^{2}=\dfrac{8\pi G}{3}\rho\,, \\  \label{friedmann1}
\dfrac{\ddot{a}}{a}=-\dfrac{4\pi G}{3}\left(\rho+3p\right)\,, \label{friedmann2} 
\end{eqnarray}
\end{subequations}
where $H\equiv\dot{a}/a$ is the Hubble rate and the absence of an index in $ \rho $ and $p$ indicates that both quantities refer to the cosmic substratum. Radiation and baryons satisfy the usual background energy conservation equation, 
\begin{subequations}
\begin{eqnarray}
\dot{\rho}_{r}+4H\rho_{r}=0\qquad&\Rightarrow&\qquad\rho_{r}=\rho_{r0}\,a^{-4}\,, \label{rhor} \\
\dot{\rho}_{b}+3H\rho_{b}=0\qquad&\Rightarrow&\qquad\rho_{b}=\rho_{b0}\,a^{-3}\,, \label{rhob}
\end{eqnarray}
\end{subequations}
where the subindex $0$ 
denotes that the corresponding quantities are evaluated at $a=a_{0}=1$.


Before treating the dark components separately, let us now combine the DM and DE components in order to describe the dark sector as an effective unified dark component. For the unified description, the total dark energy density is the sum of DM and DE energy densities whereas the total dark pressure is the DE pressure,
\begin{equation} \label{dark}
\rho_{d}=\rho_{c}+\rho_{x}\qquad{\rm and}\qquad p_{d}=p_{c}+p_{x}=p_{x}\,.
\end{equation}
From the above expression we also find
%
\begin{equation} \label{eosd}
p_{d}=\left(\frac{w_{x}}{1+r}\right)\rho_{d}\,,
\end{equation}
where $r$ is defined as the ratio between CDM and DE energy densities ($r\equiv\rho_{c}/\rho_{x}$). In the FLRW cosmology context, both functions $w_{x}$ and $r$ must depend on the scale factor, and the term in the parentheses of Eq.~\eqref{eosd} can be seen as an effective EoS parameter of the unified dark component, i.e.,
\begin{equation} \label{wd}
w_{d}\left(a\right)=\frac{w_{x}\left(a\right)}{1+r\left(a\right)}\,.
\end{equation}
Such a unified dark component must satisfy the energy conservation equation,
\begin{equation} \label{energydark}
\dot{\rho}_{d}+3H\big[1+w_{d}\left(a\right)\big]\rho_{d}=0\,.
\end{equation}
which, for a general time-dependent EoS parameter $w_{d}\left(a\right)$, has the well-known solution,
\begin{equation} \label{rhod}
\rho_{d}=\rho_{d0}\exp\left[-3\int\frac{1+w_{d}\left(a\right)}{a}da\right]\,.
\end{equation}
By splitting the total energy density in radiation, baryons and dark sector, Eq.~\eqref{friedmann1} can be rewritten as, 
\begin{equation} \label{friedmann}
3H^{2}=8\pi G\left(\rho_{r}+\rho_{b}+\rho_{d}\right)\,,
\end{equation}
where $\rho_{r}$, $\rho_{b}$, and $\rho_{d}$ are given respectively by Eqs.~\eqref{rhor}, \eqref{rhob} and \eqref{rhod}. Eqs.~\eqref{rhod} and~\eqref{friedmann} show that all information about the dark sector contained in the background expansion (Hubble rate and, consequently, any distance measurement) comes from $w_{d}(a)$. We refer the reader to Ref.~\cite{vonMarttens:2019ixw} where it was pointed out as a way to establish an explicit mapping between dynamical DE models and interacting dark sector models, so that they have identical Hubble rates. In practice, this mapping relates models that measure distances identically, and then, they can not be distinguished by distance-based observables. 

The unified EoS parameter $w_{d}$ also contains all the dark sector's contribution to the deceleration parameter 
\begin{equation} \label{deceleration}
q\equiv-\dfrac{\ddot{a}a}{\dot{a}^{2}}=\dfrac{1}{2}\sum_{i}\Omega_{i}\left(1+3w_{i}\right) \,,
\end{equation}
where $\Omega_{i}\equiv8\pi G\rho_{i}(a)/3H^{2}(a)$ is the density parameter of the $i$-th component ($i=r$, $b$ and $d$). Taking all the components into account separately, Eq.~\eqref{deceleration}  reduces to
\begin{eqnarray} \label{q}
q&=&\dfrac{1}{E^{2}}\bigg[\Omega_{r0}a^{-4}+\dfrac{1}{2}\Omega_{b0}a^{-3}+\frac{1}{2}\Big(E^{2}-\Omega_{r0}a^{-4} \nonumber \\
&& -\Omega_{b0}a^{-3}\Big)\left(1+3w_{d}\right)\bigg] \,,
\end{eqnarray}
where $E\equiv H/H_{0}$. From Eq.~\eqref{q}, it is straightforward to obtain that the acceleration condition ($q<0$) can be also formulated in terms of $w_{d}$ as,
\begin{eqnarray} \label{wdacc}
w_{d}<-\frac{1}{3}\dfrac{E^{2}+\Omega_{r0}a^{-4}}{E^{2}-\Omega_{b0}a^{-3}-\Omega_{r0}a^{-4}} \,.
\end{eqnarray}
From Eq.~\eqref{wd}, one can identify that the unified dark EoS parameter $w_{d}$ has two time-dependent degrees of freedom: $w_{x}\left(a\right)$ and $r\left(a\right)$, being each one of them related to a dynamical feature of the dark sector. The DE EoS parameter is related to the dynamical nature of the DE component, as will be seen in more detail in Sec.~\ref{sec.interacting}. Assigning a dynamic behavior to the DE component is one of the most common alternatives to the standard cosmological model \cite{Ratra:1987rm}. In this context, several parameterizations have already been proposed  for $w_{x}$ as a function of the scale factor \cite{Chevallier:2000qy,Linder:2002et,Wetterich:2004pv,Barboza:2008rh}.

On the other hand, the ratio between DM and DE energy densities $r\left(a\right)$ is associated to the existence (or not) of an interaction between the dark components. In order to understand how it is associated to a dark sector interaction, it is convenient to introduce its derivative with respect to cosmic time,
\begin{equation} \label{dr}
\dot{r}=r\left(\frac{\dot{\rho}_{c}}{\rho_{c}}-\frac{\dot{\rho}_{x}}{\rho_{x}}\right)\,.
\end{equation}
Eq.~\eqref{dr} can be combined with the conservation equations for the dark components in order to write a differential equation for the dynamics of $r\left(a\right)$. If the dark sector interacts, a source function appears in the term in the parenthesis of Eq.~\eqref{dr}, and it will directly affect the time evolution of the ratio between DM and DE energy density. The interacting case will be discussed in more detail in Sec.~\ref{sec.interacting}. 

For the $\Lambda$CDM model, the dark components are assumed to be independent, and the DE is characterized by the cosmological constant ($w_{x}=-1$) so that the background energy conservation for the dark components are given by
\begin{eqnarray}
\dot{\rho}_{c}+3H\rho_{c}=0\,, \label{rhoc} \\
\dot{\rho}_{x}=0\,. \label{rhox}
\end{eqnarray}
Substituting Eqs.~\eqref{rhoc} and \eqref{rhox} in Eq.~\eqref{dr}, one find,
\begin{equation} \label{rlcdm}
\dot{r}+3Hr=0\qquad\Rightarrow\qquad r=r_{0}\,a^{-3}\,,
\end{equation}
with the Hubble rate written as
\begin{equation} \label{Hlcdm}
H^{2}=\Omega_{r0}\,a^{-4}+\Omega_{b0}\,a^{-3}+\Omega_{d0}\frac{1+r_{0}\,a^{-3}}{1+r_{0}}\,,
\end{equation}
where $r_{0}=\rho_{c0}/\rho_{x0}=\Omega_{c0}/\Omega_{x0}$. If $w_{x}=-1$, Eq.~\eqref{rlcdm} means that any deviation from $r\propto a^{-3}$ indicates the existence of a dark sector interaction. In general, interacting models are phenomenologically proposed by an \textit{ansatz} for a source function in the energy conservation equation, but they can also be equivalently proposed by an \textit{ansatz} for the function $r\left(a\right)$ \cite{Funo:2014poa,Marttens:2016cba}. In Ref.~\cite{vonMarttens:2018bvz}, Eq.~\eqref{Hlcdm} was used to introduce a new null test sensitive to the existence of interaction in the dark sector.

\section{Interacting dark sector}
\label{sec.interacting}

As mentioned earlier, cosmologies with energy exchange between the dark sector's components constitute a viable alternative to the standard model. The observational viability of specific classes of interacting models has been investigated through the usual observational tests~\cite{Amendola:1999er,Barrow:2006hia,Clemson:2011an,Yang:2014gza,Pan:2012ki,Caprini:2016qxs,Yang:2017zjs,Yang:2018xlt,Yang:2018ubt,Yang:2019vni}, as well as through  model-independent analyses~\cite{Zhou:2019xvc}.

At the background level, being DM and DE described by perfect fluids, this non-gravitational coupling between dark components can be characterized by a scalar source term $Q$, which is the time component of the covariant derivative of the energy-momentum tensor $T^{\mu\nu}_{c\ ;\mu}=-T^{\mu\nu}_{x\ ;\mu}=\left(Q,\Vec{0}\right)$
%
%
or, equivalently, 
\begin{subequations}
\begin{eqnarray} \label{bgenergy}
\dot{\rho}_{c}+3H\rho_{c}=Q\,, \label{intrhoc} 
\end{eqnarray}
\begin{eqnarray}
\dot{\rho}_{x}+3H\rho_{x}\left(1+w_{x}\right)=-Q\,. \label{intrhox}
\end{eqnarray}
\end{subequations}
Clearly, from the above equations, the direction of the energy transfer depends on the sing of the source term: if $Q$ is positive, one finds DE decaying into CDM whereas the opposite occurs if $Q$ is negative. 

In this paper, we assume that the interaction term has the form $Q=3H\gamma R\left(\rho_{c},\rho_{x}\right)$, where $\gamma$ is a free constant  parameter and $R$ is a general function that depends on the energy densities of the components involved in the interaction. The sign of the parameter $\gamma$ gives the direction of the interaction, while its absolute value gives the strength. Note that, irrespective of the function $R\left(\rho_{c},\rho_{x}\right)$, the $\Lambda$CDM limit is always recovered for $\gamma=0$. The function $R\left(\rho_{c},\rho_{x}\right)$ has unit of energy density.

Replacing eqs.~\eqref{intrhoc} and \eqref{intrhox} into eq.~\eqref{dr}, we obtain
%
%

\begin{equation} \label{fr}
\dot{r} -3Hr\left[f\left(r\right)+w_{x}\right]=0\,,
\end{equation}
%
%
%
where
\begin{equation} \label{f}
    f\left(r\right)\equiv \gamma R\left(\dfrac{\rho_{c}+\rho_{x}}{\rho_{c}\ \rho_{x}}\right)\,.
\end{equation}
%
%
As shown in Ref.~\cite{vonMarttens:2018iav}, this approach for describing dark sector interactions is particularly interesting because, for any choice of $f\left(r\right)$, eqs.~\eqref{intrhoc} and \eqref{intrhox} can always be decoupled. Moreover, several interacting DE models proposed in the literature can be recovered. For example, the very general case $Q=3H\gamma\rho_{c}^{\alpha}\rho_{x}^{\beta}\left(\rho_{c}+\rho_{x}\right)^{\sigma}$, with $\alpha+\beta+\sigma=1$, corresponds to $f\left(r\right)=\gamma r^{\alpha-1}\left(1+r\right)^{\sigma+1}$.

Models with interaction in the dark sector are generally proposed to have a specific parameterization for the source function $Q$. However, from the definition of the function $f\left(r\right)$, it is direct to see that an equivalent approach can be	followed starting from the choice of $f\left(r\right)$. These two functions are related by
\begin{equation} \label{fQ}
Q=3H f\left(r\right)\left(\dfrac{\rho_{c}\ \rho_{x}}{\rho_{c}+\rho_{x}}\right)\,.
\end{equation}
From Eq.~\eqref{fr}, one can also show that for a given $w_{x}\left(a\right)$, the correspondence between $Q$ and $f\left(r\right)$ can be extended to $r\left(a\right)$, in the sense that any choice of $r\left(a\right)$ leads to an specific solution for $f\left(r\right)$, given by
\begin{equation} \label{fdr}
f\left(r\right)=\dfrac{a}{3r}\dfrac{dr}{da}-w_{x}\left(a\right) \,,
\end{equation}
and, consequently, to a specific solution for $Q$ provided by eq.~\eqref{fQ}. 

Tab.~\ref{tab.fqr} shows three examples for the correspondence between $f\left(r\right)$, $Q$ and $r\left(a\right)$ for the case $w_{x}=-1$. This one-to-one mapping is crucial to establish that, since there is no difference between choosing $f\left(r\right)$, $Q$ or $r\left(a\right)$ for specifying a particular interacting DE model, a model-independent reconstruction of any of these quantities provides information regarding whether or not such dark sector's interaction exists.
\begin{table}[t]
\centering
\begin{tabular}{c|c|c}\hline\hline
$f\left(r\right)$		& $Q$																				& $r\left(a\right)$ \\ \hline
$\gamma$						& $3H\gamma\dfrac{\rho_{c}\rho_{x}}{\rho_{c}+\rho_{x}}$	& $r_{0} a^{-3\left(1-\gamma\right)}$ \\ \hline
$\gamma(1+r)$					& $3H\gamma\rho_{c}$														& $\dfrac{\left(\gamma +1\right)r_{0}}{a^{3\left(\gamma+1\right)}\left(\gamma+\gamma r_{0}+1\right)-\gamma r_{0}}$ \\ \hline
$\gamma\left(1+\dfrac{1}{r}\right)$	& $3H\gamma\rho_{x}$														& $\dfrac{a^{-3\left(\gamma+1\right)} \left(-\gamma a^{3\left(\gamma+1\right)}+\gamma+\gamma r_{0}+r_{0}\right)}{\gamma+1}$ \\ \hline\hline               
\end{tabular}
\caption{Some specific cases used to illustrate the  correspondence between $f\left(r\right)$, $Q$ and $r\left(a\right)$, considering $w_{x}=-1$.}
\label{tab.fqr}
\end{table}
%
\section{Reconstructing the dark sector interaction}
\label{sec.reconst}

We shall now discuss in more detail how to reconstruct a possible dark sector interaction in a model-independent way directly from the data. To this end we address the following points in this section. First, we discuss the formalism employed, showing the explicit equations used to perform the model-independent reconstruction. Second, we present the datasets that we adopt in the analysis. Finally, we present and explore our main results.

\subsection{Set of relevant equations}
\label{ssec.eqs}

The Friedmann equation~\eqref{friedmann} can be used to write $w_{d}$ in terms of the Hubble rate. To do so, it is necessary to replace the solutions~\eqref{rhob} and~\eqref{rhod}\footnote{From now on, since we will use only low-$z$ data, the radiation component will be neglected.} in eq.~\eqref{friedmann}, and then solving the resulting equation for $w_{d}$. This leads to the following expression
\begin{equation} \label{wdz}
\small{1 + w_{d}\left(z\right)=\frac{\left(1+z\right)}{3}\frac{d}{dz}\left\lbrace\ln\left[\frac{1}{\Omega_{d0}}\left(\frac{H^{2}}{H_{0}^{2}}-\frac{\Omega_{b0}}{\left(1+z\right)^{-3}}\right)\right]\right\rbrace },
\end{equation}
which can be reduced to
\begin{equation} \label{wdz1}
    w_{d}\left(z\right)=\dfrac{E\left[3E-2\left(1+z\right)E^{\prime}\right]}{3\Omega_{b0}\left(1+z\right)^{3}-3E^{2}} \,,
\end{equation}
%
where a prime denotes derivative with respect to redshift.  From eq.~\eqref{wd}, it is direct to obtain the $\Lambda$CDM expression for $w_{d}\left(z\right)$, i.e.,
\begin{equation} \label{wdlcdm}
	w_{d}^{\Lambda{\rm CDM}}\left(z\right)=-\dfrac{\Omega_{x0}}{\Omega_{x0}+\Omega_{c0}\left(1+z\right)^{3}} \,.
\end{equation}

Eq.~\eqref{wdz1} shows explicitly how $w_{d}$ can be obtained from a reconstruction of measurements of the expansion rate. However, it is worth emphasizing that in this scenario deviations from $\Lambda$CDM result \eqref{wdlcdm} do not necessarily mean an interacting dark sector since both $w_{x}\left(z\right)$ and $r\left(z\right)$ have not been specified. In this case, deviations from the $\Lambda$CDM result may be due to a dynamic DE or an interacting dark sector. That is the essence of the so-called (background)  dark degeneracy~\cite{Kunz:2007rk,Carneiro:2014uua}.

In order to break such degeneracy, we must first specify the DE EoS $w_{x}\left(a\right)$. For simplicity, in this work we choose an interacting vacuum DE, i.e., $w_{x}=-1$. Now, combining eqs.~\eqref{wdz1} and \eqref{wd} 
the ratio between DM and DE energy densities is given by
\begin{equation} \label{rz}
	r\left(z\right)=\dfrac{-2\left(z+1\right)E E^{\prime}+6 E^{2}-3 \Omega_{b0}\left(z+1\right)^{3}}{E\left[3 E-2\left(z+1\right) E^{\prime}\right]} \,.
\end{equation} 
Therefore, any deviation from $r\left(z\right)\propto\left(1+z\right)^{3}$  means a dark sector's interaction. In this sense, as discussed earlier, the $r\left(z\right)$ reconstruction itself already has all the information on whether or not the dark sector interacts. Combining the eqs.~\eqref{wd} and \eqref{wdacc}, it is possible to write an analogous acceleration condition in terms of $r\left(z\right)$, i.e.,
\begin{equation} \label{racc}
r\left(z\right)<\dfrac{3\left(E^{2}-\Omega_{b0}a^{-3}-\Omega_{r0}a^{-4}\right)}{E^{2}-\Omega_{r0}a^{-4}}-1 \,.
\end{equation}
An equivalent procedure can be followed to obtain the DE EoS parameter given an expression of $r\left(z\right)$. However, note that a significant advantage of reconstructing $r\left(z\right)$ instead of $w_{x}\left(z\right)$ is that no assumption on the density parameters of the dark sector's components are required in the former case. On the other hand, when $w_{x}\left(z\right)$ is reconstructed, the prior information on $r\left(z\right)$ contains implicitly the value of $\Omega_{c0}$ through the relation $r_{0}=\Omega_{c0}/\left(1-\Omega_{c0}-\Omega_{b0}\right)$
~\cite{Escamilla-Rivera:2019aol}.

Finally, the last quantity we reconstruct is the interacting function $f\left(r\right)$. Combining eqs.~\eqref{fdr} and~\eqref{rz}, we obtain 
\begin{widetext}
\begin{small}
\begin{equation} \label{frz}
f\left(r(z)\right)=\frac{2\left\lbrace \Omega_{b0}\left(z+1\right)^4E\left[\left(z+1\right)E''-2E'\right]+\Omega_{b0} \left(z+1\right)^5 E^{\prime 2}-\left(z+1\right)E^{3}\left[10E'+\left(z+1\right)E''\right]+3\left(z+1\right)^{2}E^{2}E^{\prime 2}+9E^{4}\right\rbrace}{E\left[3E-2\left(z+1\right)E^{\prime}\right]\left[-2\left(z+1\right)E E^{\prime}+6E^{2}-3\Omega_{b0}\left(z+1\right)^{3}\right]} \,.
\end{equation}
\end{small}
\end{widetext}

Eqs.~\eqref{wdz1},~\eqref{rz} and~\eqref{frz} constitute the set of equations that will be reconstructed in our analysis. We do not choose to reconstruct the source term $Q$ because, according to its relation with $f\left(r\right)$, it would be necessary to solve the background energy conservation equations for the dark components (using the reconstructed solution for $f\left(r\right)$), and the error propagation would make this approach impracticable.

\subsection{Cosmological data}
\label{ssec.data}

To obtain a fully model-independent reconstruction it is not enough to use non-parametric statistical methods, but it is also necessary to use model-independent data. For this reason, we consider the following cosmological data divided into Hubble expansion rate and SN Ia luminosity distance measurements.

\subsubsection{$H(z)$ measurements}
\label{sssec:Hz}

We construct an $H(z)$ compilation with independent measurements provided by the following techniques:

\begin{itemize}
    \item \textbf{Hubble constant, $H_0$:} we adopt the cosmology-independent determination $H_0=( 75.35\pm1.68$) km/s/Mpc, which was obtained from the latest SH0ES analysis using the SN Ia distance-redshift relation calibrated via Cepheid variables~\cite{Camarena:2019moy}.
    
    \item \textbf{Cosmic Chronometers (CC):} it is possible to determine the Hubble rate of the Universe by computing the age difference $\Delta t$  between passively-evolving galaxies at close redshifts. The main requirements of the considered galaxy samples are: they have similar metallicities, low star formation rates and the average age of their stars far exceeds $\Delta t$. In the differential age method, the derivative of the cosmic time with respect to the redshift ($dt/dz$) is approximated by the ratio of the variation of galaxies' age with redshift ($\Delta t/\Delta z$). This correspondence is plausible by assuming that the analyzed galaxies were formed at the same time in the past. CC data are estimated without assuming any cosmological model and currently the 31 available  datapoints cover a wide redshift interval ($z\in [0.07, 1.965]$)~\cite{Zhang:2012mp,Simon:2004tf,Moresco:2016mzx,Stern:2009ep,Moresco:2015cya,Ratsimbazafy:2017vga}\footnote{All data points are presented in Tab.~I of the Ref.~\cite{Marra:2017pst}.}. 

    \item \textbf{Baryon Acoustic Oscillation (BAO):} we also consider the $H(z)$ estimates from the anisotropic BAO signal detected in the Luminous Red Galaxy (LRG) clustering and the quasar Ly-$\alpha$ forest. This technique does not measure directly $H(z)$, but the combination  of the Hubble rate and the scale of the sound horizon at the drag epoch, $d_{H}/r_{d}\equiv\left[ H\left(z\right)r_{d}\right]^{-1}$. In that sense, it is necessary  to calibrate all the BAO data because of their dependence on the sound horizon scale. Thus, in order to convert the combination $d_{H}\left(z\right)/r_{d}$ into a model-independent measurement of the Hubble parameter as presented in Tab.~\ref{Tab:bao}, we made use of the model-independent result for the sound horizon obtained from low-redshift standard rulers $r_{d}=\left(101.2\pm 1.8\right)h^{-1}$ Mpc~\cite{Verde:2016ccp}\footnote{We use here the value obtained in~\cite{Verde:2016ccp} that considers the condition of a spatially flat Universe.}. Combining, this result with the aforementioned local measurement of $H_{0}$~\cite{Camarena:2019moy}, the sound horizon is given by
    \begin{equation} \label{rdmi}
    r_{d}=(134.0\pm 4.0)\ \rm{Mpc}\,.
    \end{equation}

    Furthermore, to extract the 3D BAO feature from the large-scale matter distribution of the universe, it is necessary to use a fiducial cosmological model and one of the most important discussions about BAO data concerns  its dependence on this fiducial model. Recent works have shown that the constraints from BAO estimates are model-independent for a wide class of cosmological models \cite{Carter:2019ulk,Bernal:2020vbb}.

In this work, in order to avoid double counting with the BAO data used to obtain the sound horizon, we do not use the first two data points, at $z=[ 0.38, 0.51]$, provided in the full analysis of the galaxies distribution in the SDSS DR12 LRG~\cite{Alam:2016hwk}.

    %
    From the eBOSS DR16 LRG catalog~\cite{Alam:2020sor}, we use one data point at $z=0.698$. We also use the $d_{H}\left(z\right)/r_{d}$ measurement from the eBOSS DR16 QSO catalog at $z=1.48$. Finally, the analyses of the cross-correlation and the auto-correlation of the quasar Ly-$\alpha$ absorption from eBOSS DR16, that provide two uncorrelated estimates of $d_{H}\left(z\right)/r_{d}$ rate at high-$z$\footnote{We use here the Gaussian approximation of the eBOSS DR16 Ly-$\alpha$ data (auto and cross), which is a conservative approximation.}.

\begin{table}[]
\begin{tabular}{lcccc}
\hline\hline 
\multicolumn{5}{c}{BAO data}                                                                                                                                                                      \\ \hline
\multicolumn{1}{l|}{Catalog}                        & \multicolumn{1}{c|}{$z$}     & \multicolumn{1}{c|}{$H\left(z\right)$} &  \multicolumn{1}{c|}{$\sigma_{H\left(z\right)}$} & Ref.              \\ \hline
\multicolumn{1}{l|}{eBOSS DR16 LRG}                  & \multicolumn{1}{c|}{$0.698$} & \multicolumn{1}{c|}{$113.0$}                  & \multicolumn{1}{c|}{$4.16$}                           & \multirow{4}{*}{\cite{Alam:2020sor}}                  \\ \cline{1-4}
\multicolumn{1}{l|}{eBOSS DR16 QSO}                  & \multicolumn{1}{c|}{$1.48$}  & \multicolumn{1}{c|}{$170.0$}                  & \multicolumn{1}{c|}{$7.73$}                           &                  \\ \cline{1-4}
\multicolumn{1}{l|}{eBOSS DR16 Ly-$\alpha$ (cross)}  & \multicolumn{1}{c|}{$2.33$}  & \multicolumn{1}{c|}{$251.5$}                  & \multicolumn{1}{c|}{$10.63$}                           &                   \\ \cline{1-4}
\multicolumn{1}{l|}{eBOSS DR16 Ly-$\alpha$ (auto)}   & \multicolumn{1}{c|}{$2.33$}  & \multicolumn{1}{c|}{$247.0$}                  & \multicolumn{1}{c|}{$11.64$}                           &                   \\ \hline\hline 
\end{tabular}
\caption{BAO data points.}
\label{Tab:bao}
\end{table}

This calibration using one data of our set produces  correlations between  BAO-$H(z)$ measurements and $H_0$. We calculate the total correlation matrix of the $H(z)$ dataset by considering 
the BAO-$H(z)$, the sound horizon ~\eqref{rdmi} and $H_0$ values and uncertainties. The final matrix correlation of $H_0$ and BAO-$H(z)$ data is given by

\begin{equation} \label{covbao}
\begin{small}
\mathcal{C}_{H}=
\begin{bmatrix} 
2.83 & 4.25 & 6.37 & 9.38 & 9.25\\
4.25 & 17.32 & 15.59 & 23.01 & 22.66\\
6.37 & 15.59 & 59.81 & 34.53 & 33.96\\
9.38 & 23.01 & 34.53 & 112.95 & 50.24\\
9.25 & 22.66 & 33.96 & 50.24 & 135.44\\
\end{bmatrix} ,
\end{small}
\end{equation}

\noindent where the first row and column correspond to the $H_0$ correlation coefficients. Note that the CC data is uncorrelated, thus its correspondent block in the total covariance matrix is diagonal.

 \end{itemize}


\subsubsection{Type Ia Supernovae}

The second dataset we use is the Pantheon catalog of type Ia Supernovae (SNe Ia), which contains 1048 data points in the range $0.01<z<2.26$, with their respective covariance matrix (including statistical and systematic errors)\footnote{The data as well as its covariance matrix can be found at \href{https://github.com/dscolnic/Pantheon}{github.com/dscolnic/Pantheon}}. The data obtained from SN Ia is not directly the Hubble rate, but the $B$-band apparent magnitude $m_{B}$, which is related to the luminosity distance $d_{L}\left(z\right)$ by
\begin{equation} \label{distmodulus}
m_{B}=5\log_{10}\left[\dfrac{d_{L}\left(z\right)}{1{\rm Mpc}}\right]+25+M_{B}\,,
\end{equation}
where $M_{B}$ is the $B$-band absolute magnitude of the SNe~Ia. 

We directly relate $d_L$ with the comoving distance ($d_c$) by considering a spatially flat universe via:

\begin{equation}
d_c(z)=\frac{d_L(z)}{1+z}=\frac{10^{(m_B-M_B-25)/5}}{1+z}.
\label{Dc}
\end{equation}
Thus, to convert this $m_{B}$ SNe Ia catalog into comoving-distance data it is just necessary to know the absolute magnitude value. In general, for parameter selection analyses, $M_{B}$ is considered as a nuisance parameter. Here, however, we fix it at the value~\cite{Camarena:2019moy}
\begin{equation} \label{Mmi}
M=-19.2334\pm 0.0404\, {\rm mag}\,.
\end{equation}
The previous approach allows us to directly calculate the Hubble rate function with the reconstruction of the first derivative of $d_c$ through its definition,
\begin{equation} \label{D}
{d_c}\left(z\right)=\int_{0}^{z}\dfrac{d\tilde{z}}{H\left(\tilde{z}\right)}\quad\Rightarrow\quad H\left(z\right)=\dfrac{1}{d_c^{\prime}\left(z\right)}\,.
\end{equation}
%

It is worth to highlight that although the $M_B$ considered is also model independent, this value is not relevant in our analysis because all the interaction quantities depend only on the normalized Hubble parameter ($E\equiv H(z)/H_0=d_{c}'(0)/d_{c}'(z)$) and therefore the multiplicative factor $10^{-M/5}$ in Eq.~\eqref{Dc} cancels. In this way, the SNe data becomes almost model independent.

\begin{figure}[t]
\includegraphics[width=\columnwidth, trim={1cm 1.3cm 1cm 1cm}, clip]{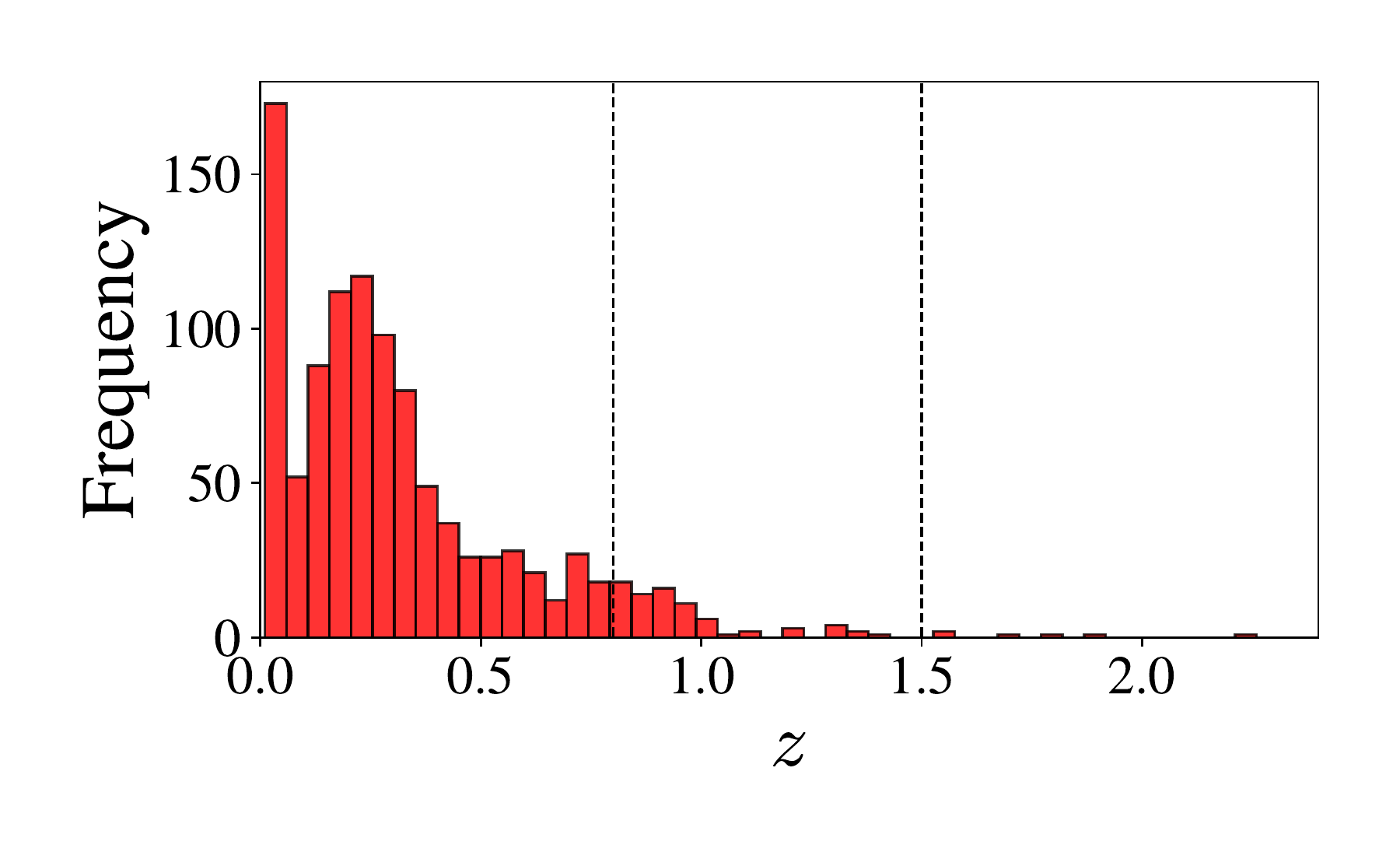}
\caption{Distribution of the SN Ia data points in terms of the redshift. The dashed lines indicate $z=0.8$ and $z=1.5$, dividing the redshift range into three intervals. The first interval contains $92.18\%$ of the Pantheon's catalog data (966 SNe Ia), the second interval includes $7.25\%$ (76 SNe Ia), remaining only $0.57\%$ (6 SNe Ia) in the last interval.}
\label{histsn}
\end{figure}

Another important aspect in our SN Ia analysis is that the SN Ia data distribution along the redshift is not uniform. In fact, there are many more SNe Ia at low-$z$ than at high-$z$, as shown in Fig.~\ref{histsn}. For this reason, we show only the data set within the redshift range used in our analysis. 
The redshift threshold for the SN Ia reconstruction will be discussed on a case-by-case basis in Sec.~\ref{ssec.results}.
\begin{figure*}[t]
\includegraphics[width=\columnwidth, trim={0 1cm 0 0,7cm}, clip]{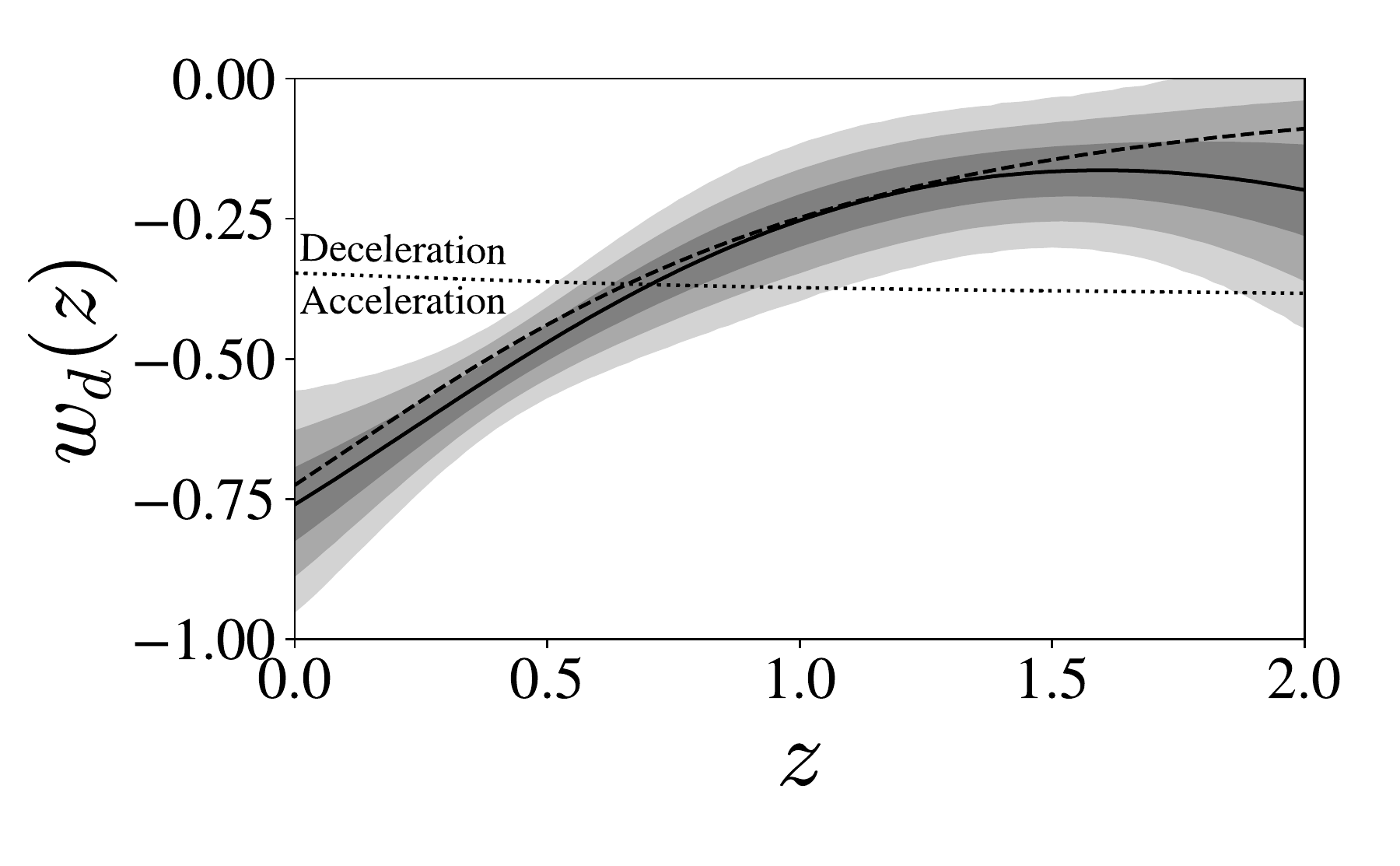}
\includegraphics[width=\columnwidth, trim={0 1cm 0 0,7cm}, clip]{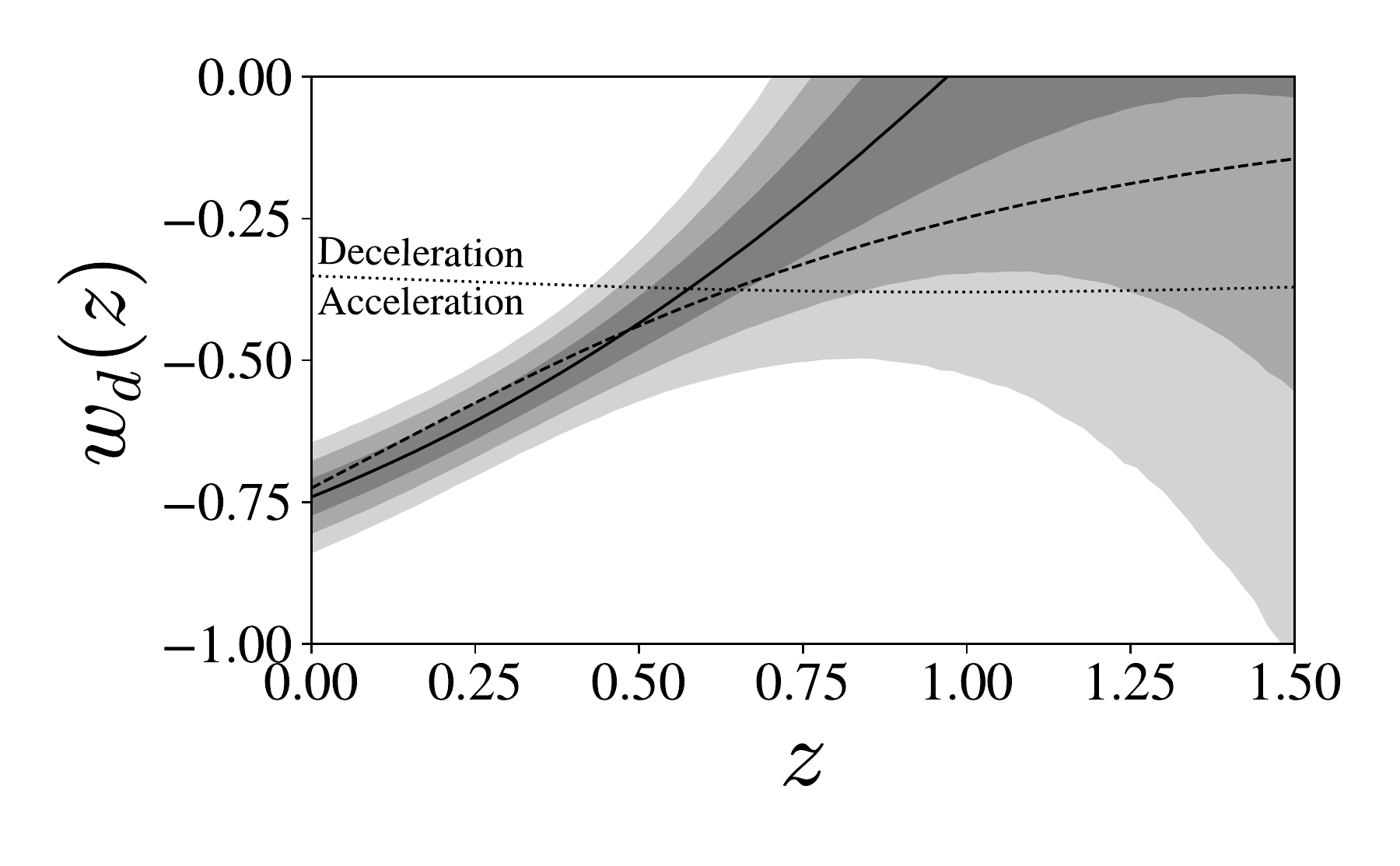}
\caption{Model-independent reconstruction of the EoS parameter of the unified dark fluid. The solid line is the best fit of the GP reconstruction, the dashed line is the $\Lambda$CDM result, and the dotted line is the acceleration condition, given by Eq.~\eqref{wdacc}. \textbf{Left panel: }Result obtained using the $H(z)$ data. \textbf{Right panel: }Result obtained using the SN Ia data.}
\label{figwd}
\end{figure*}

\subsubsection{Reconstruction}

The model-independent reconstruction is performed by applying the Gaussian Process (GP) method. We use the GaPP python library \cite{Seikel:2012uu} with a square exponential covariance function and optimize its hyperparameters by maximizing the GP's likelihood to obtain the reconstruction of $H(z)$, $d_c(z)$ and their derivatives. As mentioned earlier, the quantities that we will reconstruct are $w_d(z)$, $r(z)$ and $f(z)$, which are respectively given by Eqs.~\eqref{wdz1},~\eqref{rz} and~\eqref{frz}. For the first data set, we extract the Hubble rate directly from the data. The normalization of the Hubble rate is made considering the value of $H\left(z=0\right)=H_{0}$ obtained with the reconstruction. 
On the other hand, for the second data set, the reconstructed quantity is initially the $d_c$, and the (normalized) Hubble rate is obtained following  Eq. ~\eqref{D}. Note that, compared with the first data set, the SN Ia analysis will always need one more derivative, which may affect the error propagation. 

In order to reconstruct $w_d(z)$, $r(z)$ and $f(z)$ (Eqs.~\eqref{wdz1},~\eqref{rz} and~\eqref{frz}),  we perform a Monte Carlo sampling taking into account the mean values and the complete covariance  matrix of $H_0$, $H(z)$, $H^{\prime}(z)$ and $H^{\prime\prime}(z)$ jointly described by a GP (multivariate Gaussian distribution), for the first data set. In the case of SN data,  we perform the Monte Carlo sampling with $d^{\prime}_{c}(0)$, $d^{\prime}_{c}(z)$, $d^{\prime\prime}_{c}(z)$ and $d^{\prime\prime\prime}_{c}(z)$ mean values and their covariance matrix (see Eq. (2.6) of Ref. \cite{Seikel:2012uu}). Another prior information required by the Monte Carlo sampling of eqs.~\eqref{wdz1},~\eqref{rz} and~\eqref{frz} is the baryon density parameter $\Omega_{b0}$. Here, we use a result obtained in Ref.~\cite{Pettini:2012ph},  $\omega_{b}\equiv\Omega_{b0}h^{2}=0.0223\pm 0.0009$. When combined with the local measurement of $H_{0}$ presented in Sec.~\ref{sssec:Hz}, we obtain $\Omega_{b0}=0.0393 \pm 0.0024$.

\begin{figure*}[t]
\includegraphics[width=\columnwidth, trim={0 1cm 0 0,7cm}, clip]{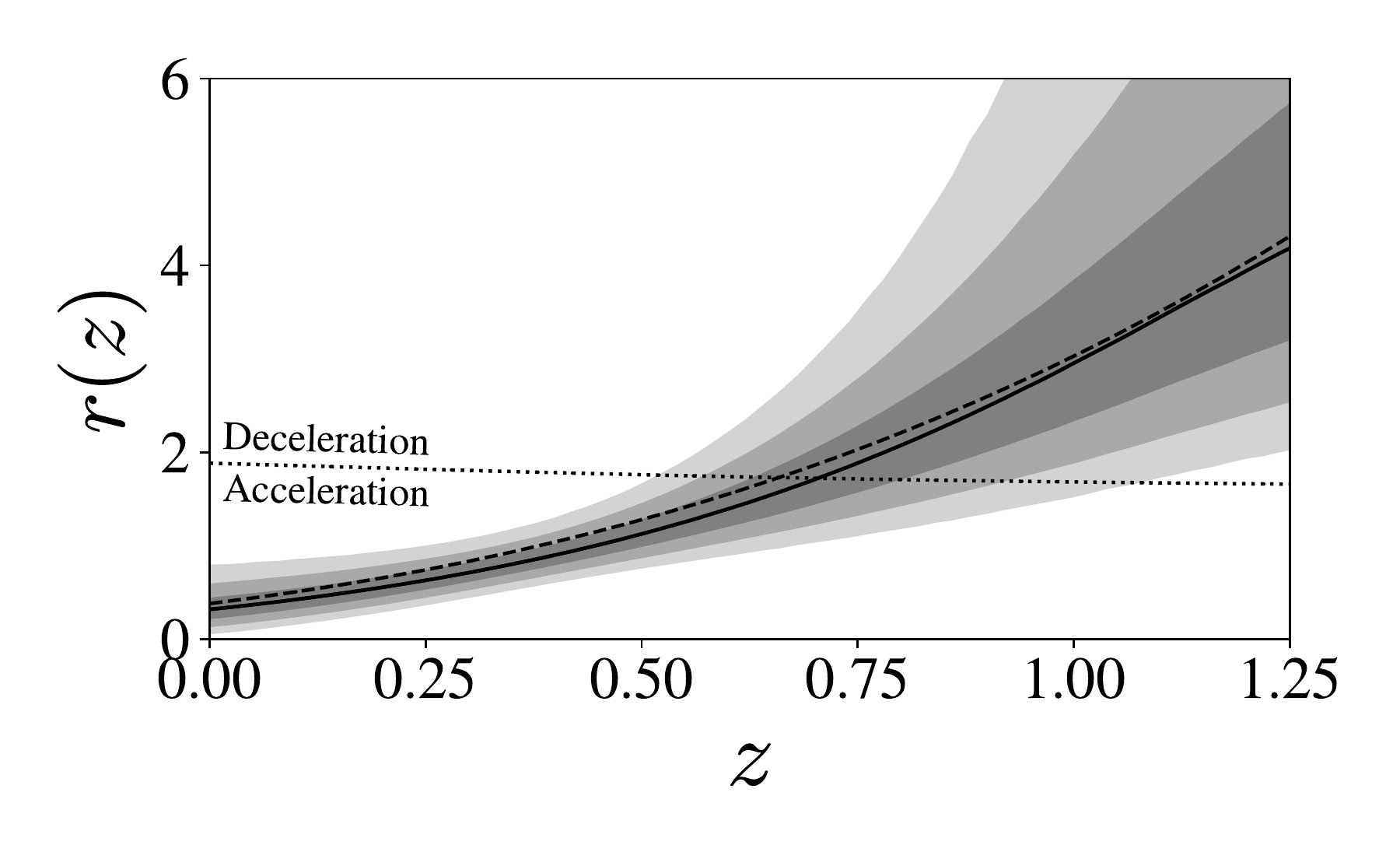}
\includegraphics[width=\columnwidth, trim={0 1cm 0 0,7cm}, clip]{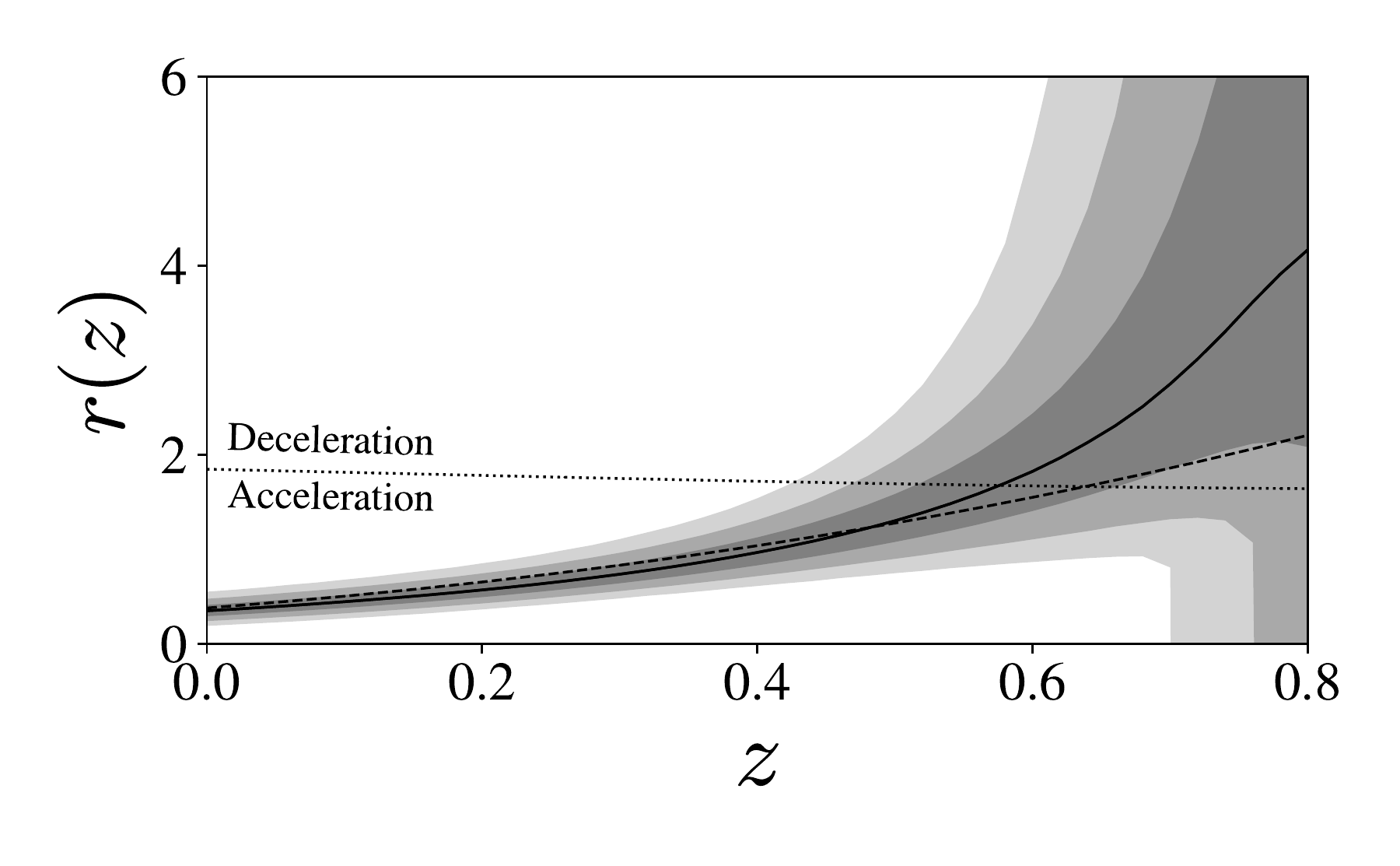}
\caption{Model-independent reconstruction of the ratio between CDM and DE energy densities. The solid line is the best fit of the GP reconstruction, the dashed line is the $\Lambda$CDM result, and the dotted line is the acceleration condition, given by Eq.~\eqref{racc}. \textbf{Left panel: }Result obtained using the $H(z)$ data. \textbf{Right panel: }Result obtained using the SN Ia data.}
\label{figrz}
\end{figure*}
%

\subsection{Results}
\label{ssec.results}

In this section we present and discuss our results for the reconstruction of the possible interaction in the cosmological dark sector. The first quantity  reconstructed is the unified dark EoS parameter, given by Eq.~\eqref{wdz1}. As already mentioned, deviations from $w_d(z)=w_{d}^{\Lambda{\rm CDM}}\left(z\right)$ do not necessarily mean a dark sector interaction, but can also represent a dynamical DE or even a combination of both. 
\begin{figure*}[t]
\includegraphics[width=\columnwidth, trim={0 1cm 0 0,7cm}, clip]{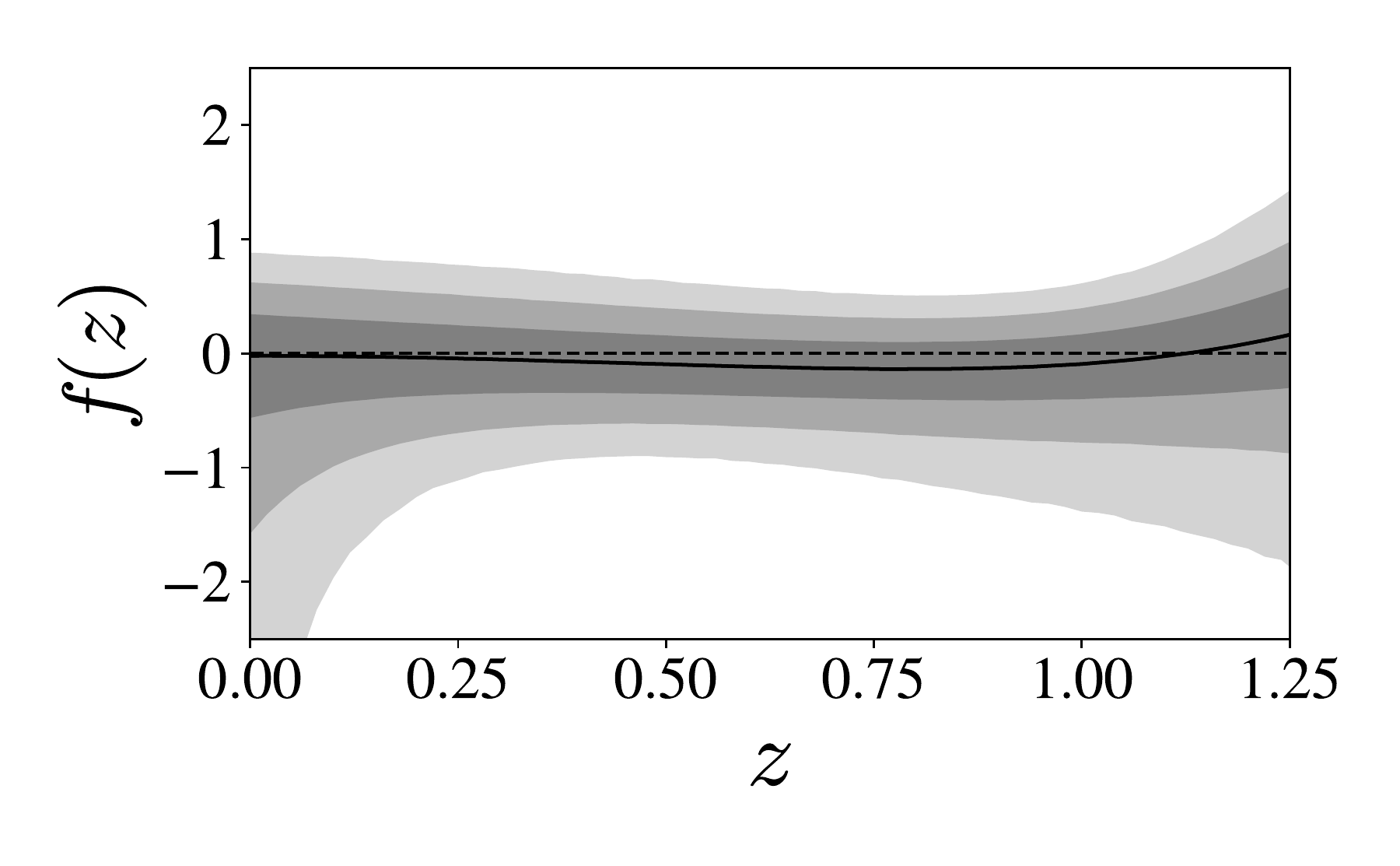}
\includegraphics[width=\columnwidth, trim={0 1cm 0 0,7cm}, clip]{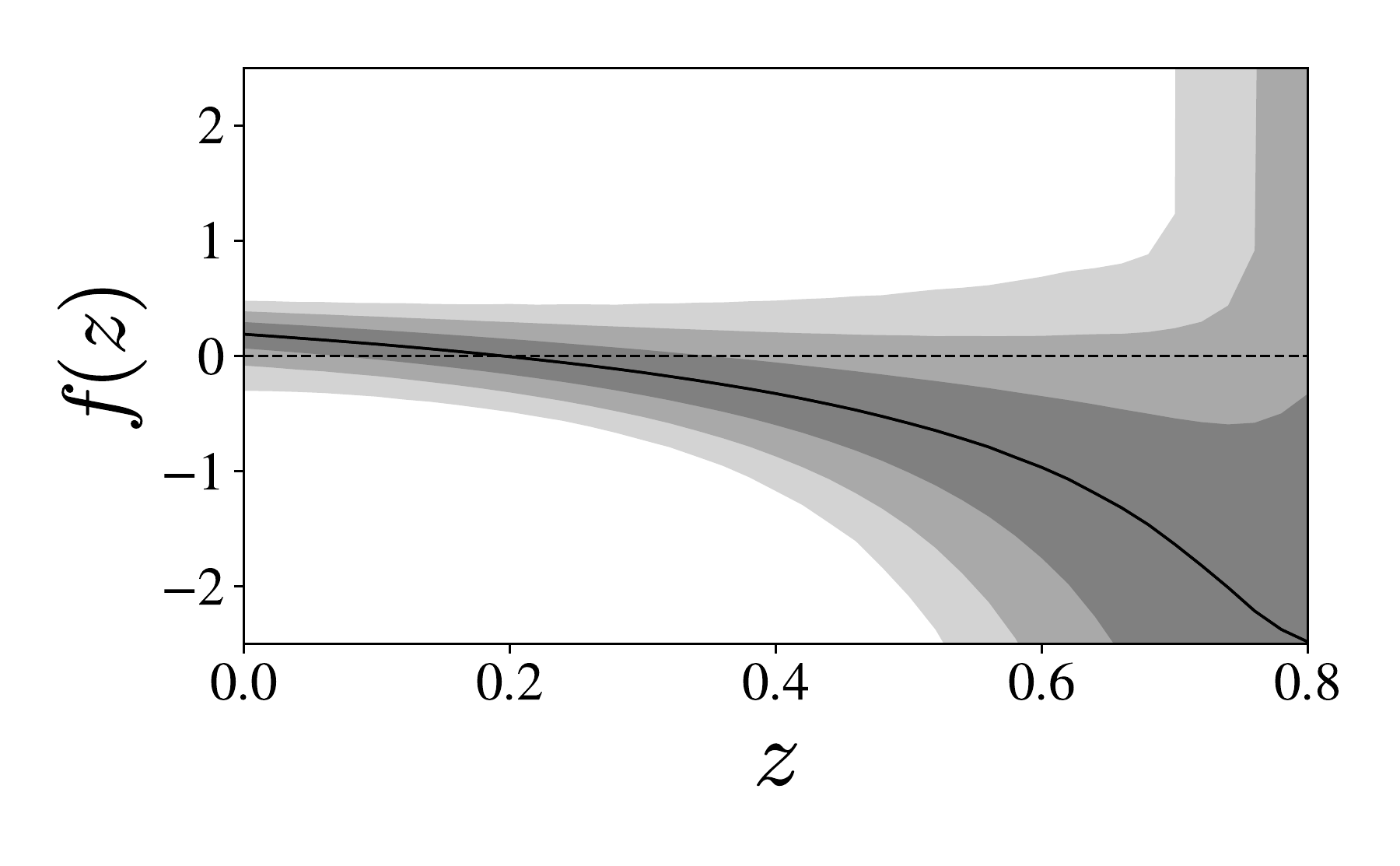}
\caption{Model-independent reconstruction of the dark sector interaction function in terms of the redshift. The dashed horizontal line in zero, indicates the $\Lambda$CDM (non-interacting) result. \textbf{Left panel: }Result obtained using the $H(z)$ data. \textbf{Right panel: }Result obtained using the SN Ia data.}
\label{figfz}
\end{figure*}
Fig.~\ref{figwd} shows the $w_{d}\left(z\right)$ reconstruction. Whereas its left panel shows the results obtained using $H (z)$ measurements, the right panel shows the results from SN Ia data. In both cases, one can see that $\Lambda$CDM model\footnote{From now on, when $\Lambda$CDM model is compared to our results, it is implicit that we refer to the best fit of the Planck analysis using data from TT,TE,EE+lowE+lensing+BAO \cite{Aghanim:2018eyx}.} is consistent at least at $3\sigma$ CL over the entire redshift range.  Furthermore, in both cases, the best fits of the reconstructions indicate that the universe switched from a decelerated phase to an accelerated one at around $z\approx 0.6$. We estimate the CL of $w_d$ at $z=0$ up to $5\sigma$ via Monte Carlo sampling for the two data sets and confirm the present cosmic acceleration with  a higher CL. Our results indicate current acceleration  at $\sim 5.9\sigma$ CL with the $H(z)$ data and, if we assume the gaussianity is maintained at higher levels, the SNe Ia data confirms cosmic acceleration at $\sim 12\sigma$ CL.
It is worth emphasizing that these results only assume a dark sector composed of a pressureless matter (whether interacting or not) combined with a general DE component, which is utterly free from a hypothesis about its nature. In the SNe Ia analysis, even though all data points have been used, we restrict ourselves to show only the result for $z<1.5$. For higher values of the redshift, the SNe Ia analysis can not properly constraint $w_{d}(z)$, and it has no physical meaning.

Imposing now $w=-1$, we reconstruct the ratio between DM and DE energy densities, given by Eq.~\eqref{rz}. Now any deviation from $r\left(z\right)\propto\left(1+z\right)^{3}$ indicates the existence of a non-gravitational interaction between the dark components. The results for the $H\left(z\right)$ and SN Ia data sets are presented respectively in the left and right panels of Fig.~\ref{figrz}. As well as the first case, both  analyses have an excellent agreement with the $\Lambda$CDM model, so that the reconstruction is consistent in $1\sigma$ CL with the standard cosmological model across the whole redshift interval analyzed. Once again, as the first analysis, the best fits of the $r\left(z\right)$ reconstructions predict a transition from a decelerated phase to an accelerated phase at $z\approx 0.6$. Using Eq.~\eqref{racc}, the cosmic acceleration can be assessed in terms of $r(z)$. The evidence that universe now ($z=0$) experiences an accelerated expansion remains at more than $6\sigma$ for both data sets. Even though this analysis considers only deviation associated with dark sector interactions, similar conclusions in comparison to the $w_{d}(z)$ analysis are indeed expected because of the dark degeneracy. Here, the SN Ia analysis only delivers reasonable constraints for $z<0.8$.
\begin{figure}[t!]
\centering
\includegraphics[width=\columnwidth, trim={0 2.84cm 0 0,7cm}, clip]{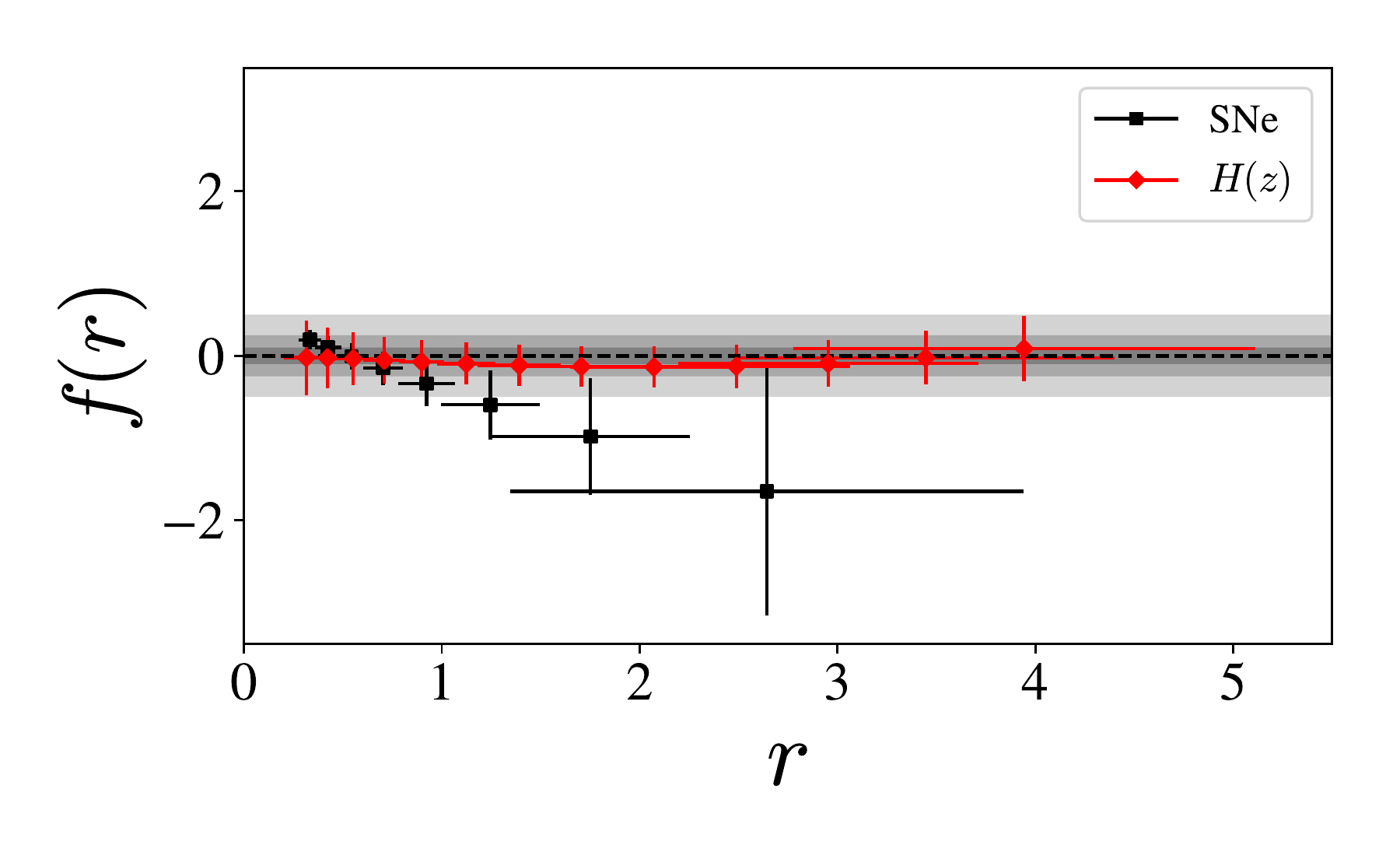}\vspace*{-0.5mm}
\includegraphics[width=\columnwidth, trim={0 2.84cm 0 0.9cm}, clip]{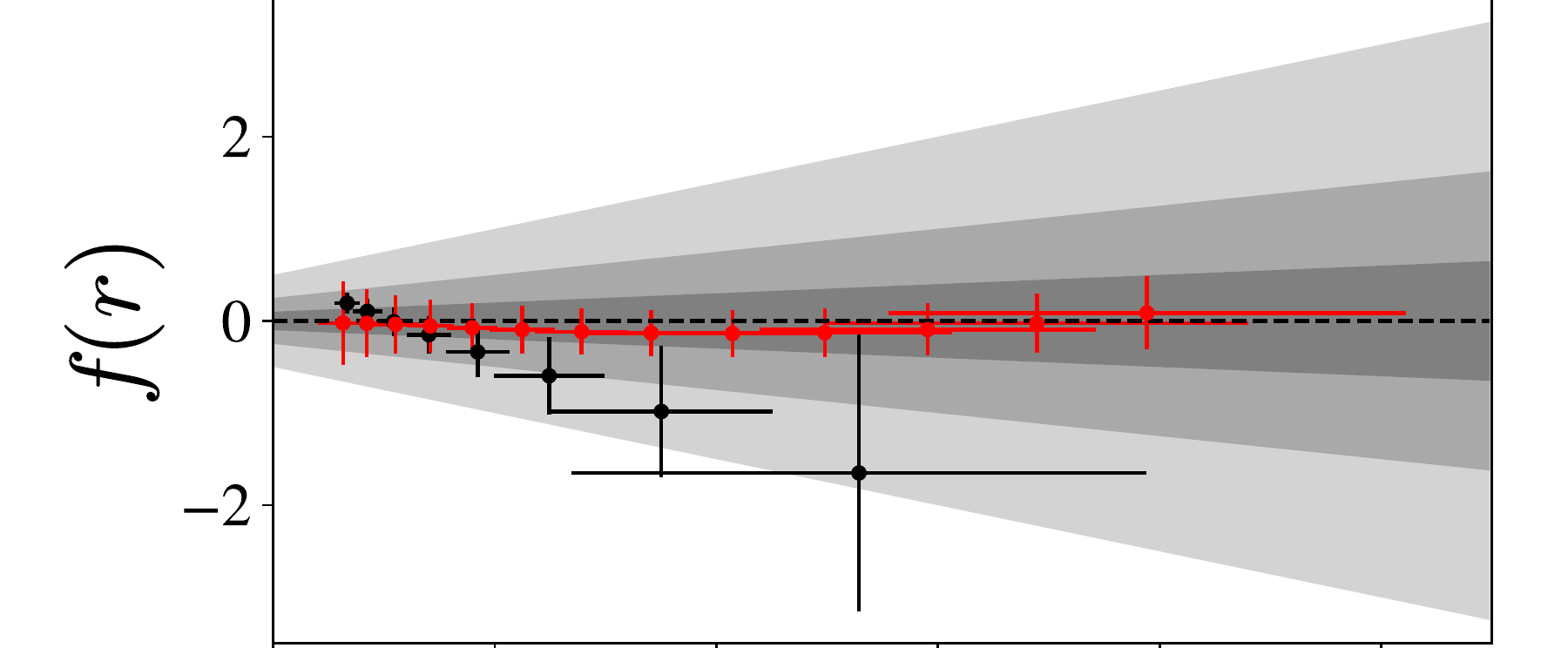}\vspace*{-0,5mm}
\includegraphics[width=\columnwidth, trim={0 1cm 0 0,9cm}, clip]{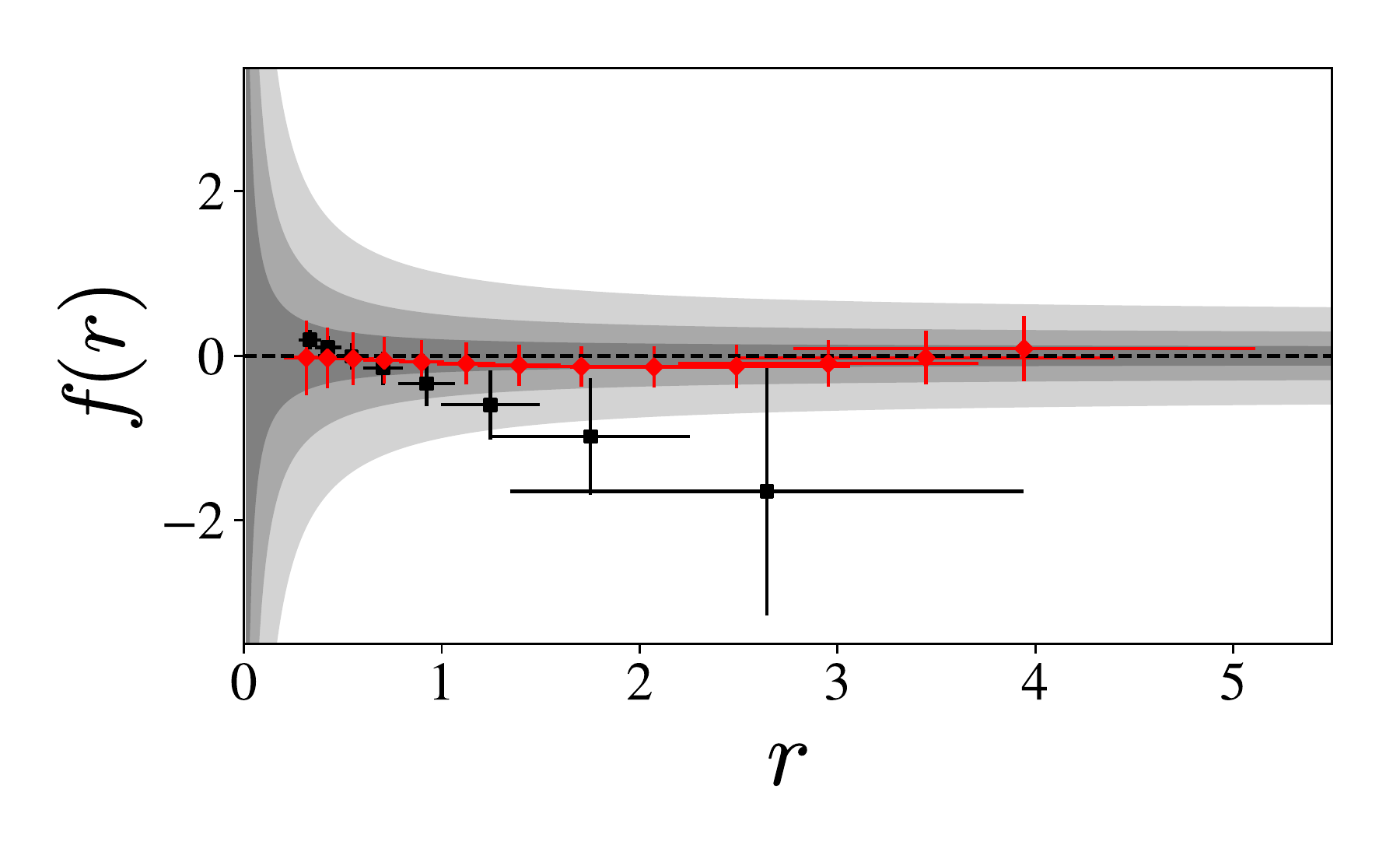}
\caption{Parametric combination of the model-independent reconstructions for the interacting function $f$ and the ratio between CDM and DE energy densities $r$. In each panel, the result is compared with one of the interacting models shown in Tab.~\ref{tab.fqr} with $\gamma=\pm0.1,\pm0.25,\pm0.5$ corresponding to bands in dim gray, standard grey and light grey respectively. \textbf{Top panel: }$f\left(r\right)=\gamma$. \textbf{Middle panel: }$f\left(r\right)=\gamma(1+r)$. \textbf{Bottom panel: }$f\left(r\right)=\gamma(1+1/r)$.}
\label{figfr}
\end{figure}

The last reconstruction we perform is for the  interacting function, defined in Eq.~\eqref{f}, in terms of the redshift, which is given by Eq.~\eqref{frz}. As already presented in Tab.~\ref{tab.fqr}, this quantity identifies uniquely in a simple way the source term $Q$, and for this reason, its reconstruction can be seen as a model-independent reconstruction of the dark sector interaction. The results using $H(z)$ and SN Ia data are shown respectively in the left and right panels of Fig.~\ref{figfz}. In both analysis the $\Lambda$CDM model agrees within $2\sigma$ CL for the entire redshift range. 

Finally, since the relation between the interaction function and the source term is more evident when $f$ is written in terms of the ratio between CDM and DE energy densities, we take advantage of the $r(z)$ and $f(z)$ results to perform a parametric investigation of $f(r)$. In Fig.~\ref{figfr}, we show the $1\sigma$ CL results for the interacting function in terms of the ratio between CDM and DE energy densities, being the red diamonds the results obtained with the $H(z)$ measurements and the black squares obtained with the SN Ia data. In each of the three panels we compare our results with the models highlighted in Tab.~\ref{tab.fqr}, where the regions between $\gamma=\pm0.1$, $\gamma=\pm0.25$ and $\gamma=\pm0.5$ are respectively associated to dim gray, standard grey and light grey. The top, middle and bottom panels of Fig.~\ref{figfr} show the result for the cases $f(r)=\gamma$, $f(r)=\gamma (1+r)$ and $f(r)=\gamma (1+1/r)$. Our results confirm that the $\Lambda$CDM model (dashed line in $f(r)=0$) is compatible with our results in about $1\sigma$ CL, but also indicate that current data has no strength  to discard the interacting models.

\section{Future perspectives}
\label{sec.futperspec}

\begin{figure}
\includegraphics[width=\columnwidth, trim={0 1.3cm 0 0,6cm}, clip]{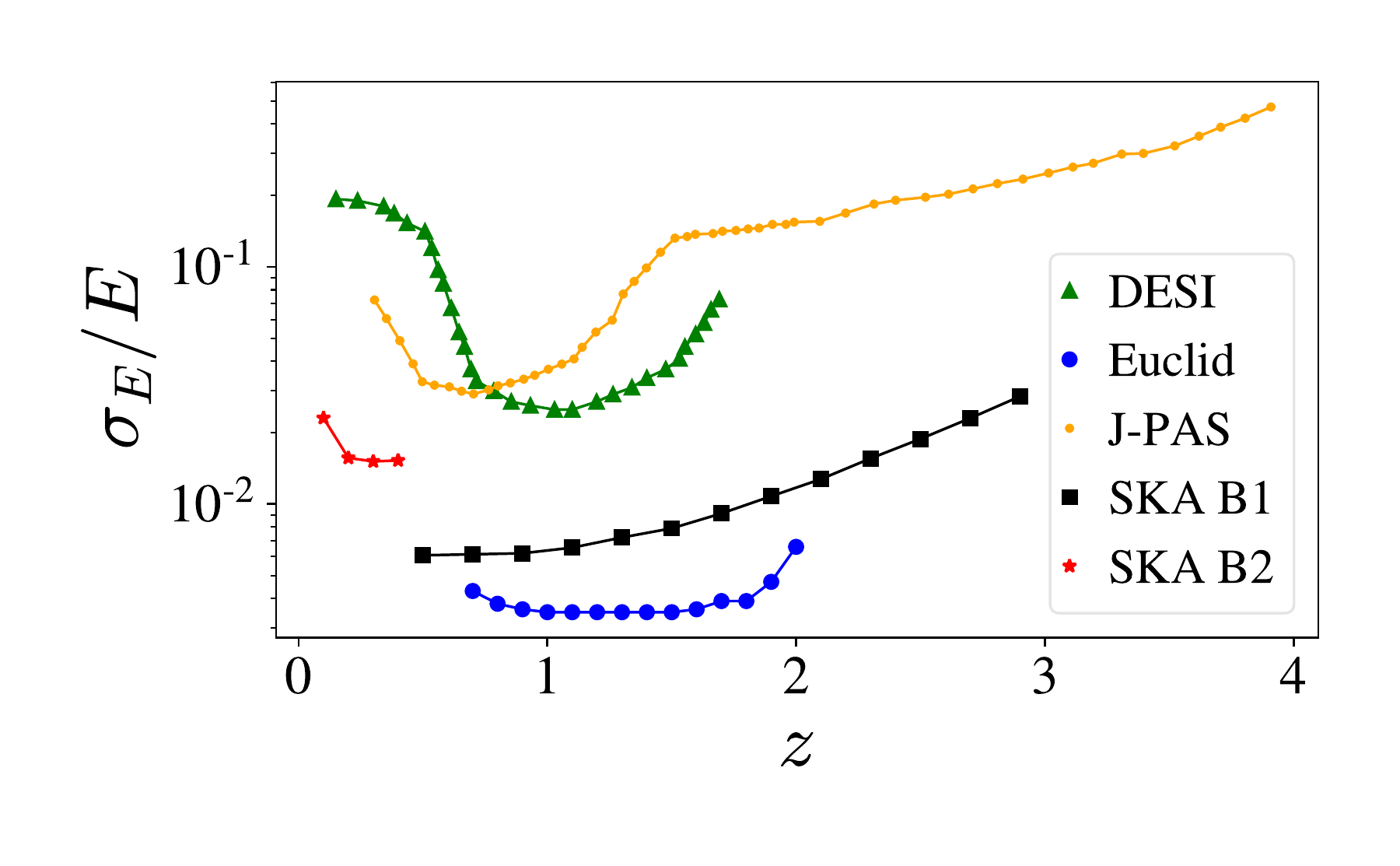}
\caption{Forecasted relative error on $\sigma_{E/E}$.}
\label{fig.forecastdata}
\end{figure}

With the advent of the next-generation LSS surveys, the subsequent few years promise to be particularly fruitful for observational cosmology. Among others, we can highlight J-PAS~\cite{Benitez:2014ibt,Bonoli:2020ciz}, DESI~\cite{Aghamousa:2016zmz}, EUCLID~\cite{Amendola:2012ys} and SKA~\cite{Bacon:2018dui}. 
These surveys will provide measurements of $H\left(z\right)$ from BAO with good precision in a redshift interval where there are not many SNe Ia ($z \gtrsim 0.8$). In that sense, this data will play a crucial role as a complementary data for unveiling the nature of the dark sector of the Universe. In this work, we use the forecast for the (normalized) Hubble rate from J-PAS, DESI, EUCLID and SKA (bands 1 and 2) presented in~\cite{Benitez:2014ibt,Bacon:2018dui}. The specifications of the catalogs we use are the following:
\begin{flalign}
  &\qquad \bullet\ {\rm J-PAS:} & 0.31<&\ z<3.91 &,\qquad &N=54  \\[0.2cm]
  &\qquad \bullet\ {\rm DESI:} & 0.15<&\ z<1.691 &,\qquad &N=30  \\[0.2cm]
  &\qquad \bullet\ {\rm Euclid:} & 0.7<&\ z<2.0 &,\qquad &N=14 \\[0.2cm]
  &\qquad \bullet\ {\rm SKA\ B1:} & 0.5<&\ z<2.9 &,\qquad &N=12 \\[0.2cm]
  &\qquad \bullet\ {\rm SKA\ B2:} & 0.1<&\ z<0.4 &,\qquad &N=4
\end{flalign}
where $N$ is the number of points in each survey. The results of the relative error $\sigma_{E/E}$ are shown in Fig.~\ref{fig.forecastdata}. As can be seen, the surveys will be able to provide a very important complementary data to investigate the DE component and the late-time cosmic expansion. In particular, we find that Euclid and SKA (band 1) will deliver sub-percent measurements of $H\left(z\right)$ at $0.5<z<2$. We also confirm the very good performance of J-PAS at low-$z$ ($z \lesssim 0.7$), as reported in ~\cite{Resco:2019xve}.

\begin{figure}
\includegraphics[width=\columnwidth, trim={0 2,87cm 0 0}, clip]{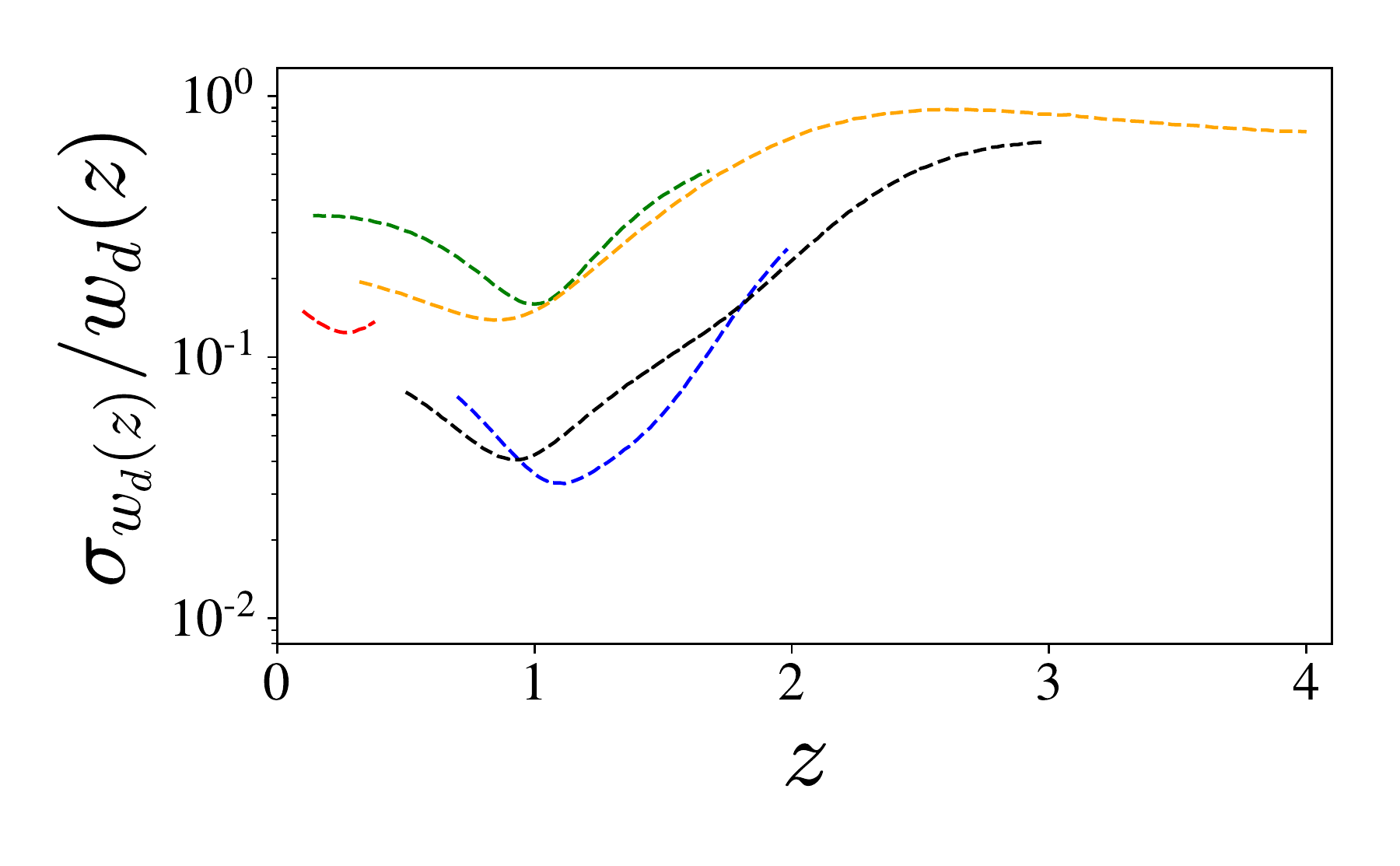}\vspace*{-0,4mm}
\includegraphics[width=\columnwidth, trim={0 1cm 0 0.9cm}, clip]{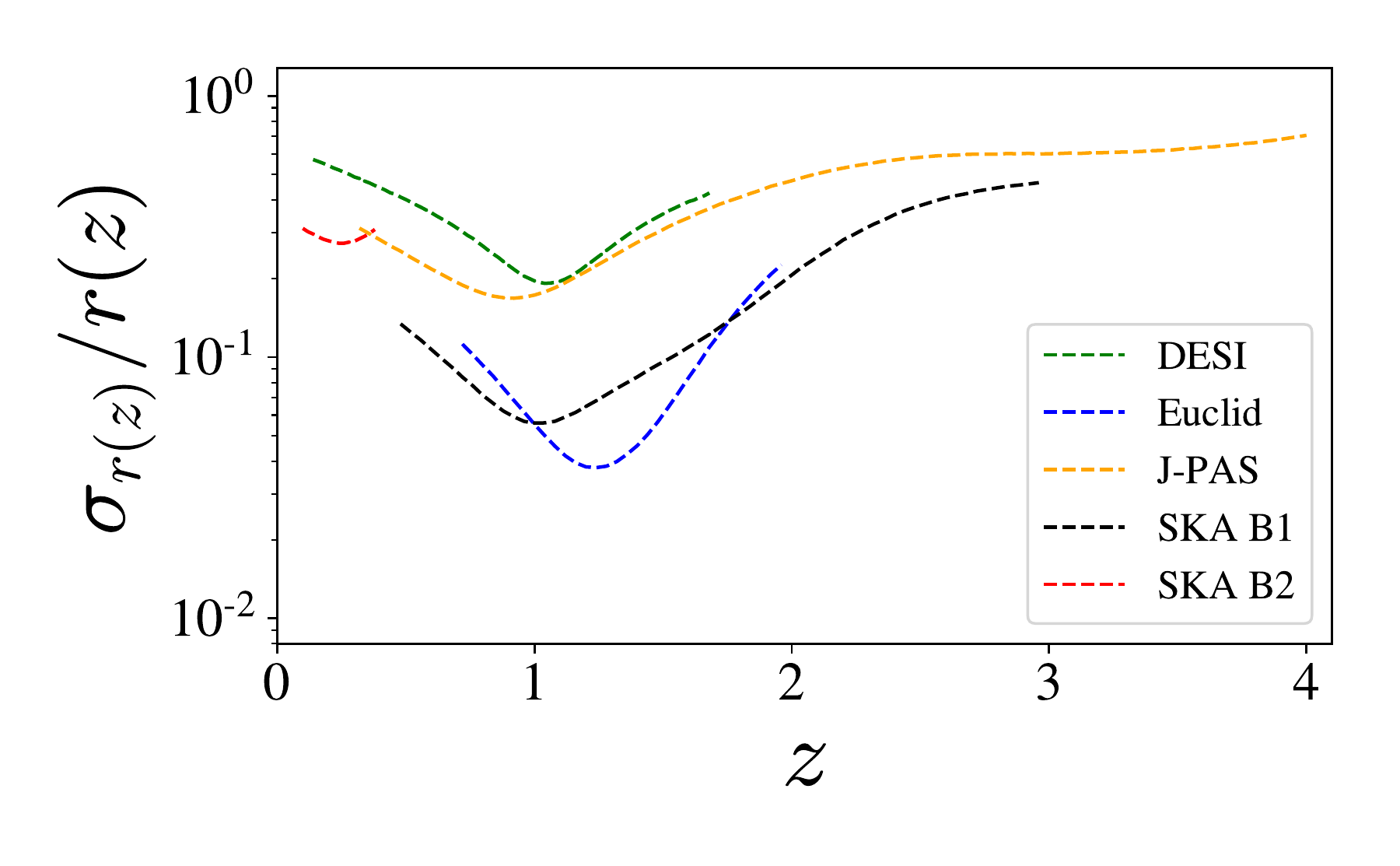}
\caption{Results of the forecast analysis. \textbf{Top panel:} Forecasted relative error on $\sigma_{w_{d}\left(z\right)}/w_{d}\left(z\right)$. \textbf{Bottom panel:} Forecasted relative error on $\sigma_{r\left(z\right)}/r\left(z\right)$.}
\label{fig.forecastresult}
\end{figure}

Our forecast analysis is similar to one presented in~\cite{Bengaly:2019oxx,Bengaly:2019ibu,Bengaly:2020neu}. Using the data presented in Fig.~\ref{fig.forecastdata}, we build a $\Lambda$CDM ($\Omega_{m}=0.3$\footnote{Note that, since only the normalized Hubble parameter is used, no assumption for $H_{0}$ is required.}) mock catalog for each survey, and {{use GP to reconstruct them}}. First, we apply the reconstructions of the (normalized) Hubble rate in Eq.~\eqref{wdz1}. In this case, we forecast the EoS parameter for the dark sector as a whole, i.e., no assumption on the time-dependence of the DE component, or whether it interacts or not with CDM is made. The relative error on $\sigma_{w_{d}\left(z\right)}/w_{d}\left(z\right)$ is presented in the top panel of Fig.~\ref{fig.forecastresult}. Finally, assuming that DE is described by $w_{x}=-1$, but admitting the possibility of an interaction in the dark sector, we apply the reconstructed (normalized) Hubble rate in Eq.~\eqref{rz}. In this case, we forecast the ratio between CDM and DE energy densities, which, for a given $w_{x}\left(a\right)$, determines uniquely an interaction between the dark components. In that sense, this analysis shows the strength of the next-generation LSS surveys for constraining interacting models in a model-independent way. The result for the relative error on $\sigma_{r\left(z\right)}/r\left(z\right)$ is presented in the bottom panel of Fig.~\ref{fig.forecastresult}. In both forecast analyses, we highlight that Euclid and SKA (band 1) will be able to constraint $w_{d}\left(z\right)$ and $r\left(z\right)$ with $\sim4\%$ of precision at $z\approx 1$.
%

\section{Conclusions}
\label{sec.conclusions}

A number of recent analysis have shown that models with interaction in the dark sector are able to provide a good description of the current observational data, which makes them a promising alternative to the standard cosmology. However, the absence of guidance from a fundamental theory on the coupling term $Q$ makes its choice arbitrary. 

In this paper we tested the possibility of a non-minimal coupling between the CDM and DE components by reconstructing the functions $w_d(z)$, $r(z)$ and $f(r)$, as defined in Secs.~\ref{sec.dark}, and~\ref{sec.interacting}. Using currently available data of type Ia Supernova, Cosmic Chronometers and Baryonic Acoustic Oscillations, we followed the model-independent approach discussed in Sec.~\ref{sec.reconst} and showed  that the current observations show a good agreement (within 3$\sigma$ level) with the $\Lambda$CDM hypothesis of uncoupled dark components ($Q = 0$), although an interaction between them cannot be completely ruled out. For completeness, we also performed a forecast analysis for the next-generation of large-scale structure surveys considering their predictions for the normalized expansion rate $E(z)$. We found that J-PAS will have a better performance at low-$z$ ($\lesssim 0.7$) when compared with DESI and that Euclid and SKA (band 1) will be able to constrain the functions $w_{d}\left(z\right)$ and $r\left(z\right)$ with $\sim4\%$ of precision at $z\approx 1$. Such results clearly show that the upcoming data from the next-generation surveys will play a crucial role in probing the possibility of a non-minimal coupling between the components of the cosmological dark sector.


\section*{Acknowledgements}
It is a pleasure to thank Carlos A. P. Bengaly for sharing the data used in the forecast analysis.
RvM acknowledges support from the Programa de Capacitação Institucional PCI/ON/MCTI. JA acknowledges support from CNPq (Grants no. 310790/2014-0 and 400471/2014-0) and FAPERJ (grant no. 233906).
JEG, VM and LC thanks CNPq (Brazil) for partial financial support. This project has received funding from the European Union’s Horizon 2020 research and innovation programme under the Marie Skłodowska-Curie grant agreement No 888258.

\bibliography{rz}

\end{document}